\documentclass[aps,prl,twocolumn,showpacs,preprintnumbers]{revtex4}
\usepackage{graphicx}
\usepackage{dcolumn}
\usepackage{bm}

\def\be{\begin{equation}}
\def\ee{\end{equation}}
\def\bea{\begin{eqnarray}}
\def\eea{\end{eqnarray}}
\newcommand{\beq}{\begin{equation}}
\newcommand{\eeq}{\end{equation}}

\newcommand{\rc}{\nonumber\\}
\newcommand{\bear}{\begin{eqnarray}}
\newcommand{\eear}{\end{eqnarray}}

\def\lbldef#1#2{\expandafter\gdef\csname #1\endcsname {#2}}

\def\href#1#2{#2}

\newcommand{\ber}{\begin{eqnarray}}
\newcommand{\eer}{\end{eqnarray}}

\newcommand{\beqar}{\begin{eqnarray}}

\newcommand{\eeqar}{\end{eqnarray}}

\newcommand{\dsl}
  {\kern.06em\hbox{\raise.15ex\hbox{$/$}\kern-.56em\hbox{$\partial$}}}

\newcommand{\eeqarr}{\end{eqnarray}}
\newcommand{\ZZ}{{\rm \kern 0.275em Z \kern -0.92em Z}\;}

\def\CC{{\mathchoice
{\rm C\mkern-8mu\vrule height1.45ex depth-.05ex
width.05em\mkern9mu\kern-.05em}
{\rm C\mkern-8mu\vrule height1.45ex depth-.05ex
width.05em\mkern9mu\kern-.05em}
{\rm C\mkern-8mu\vrule height1ex depth-.07ex
width.035em\mkern9mu\kern-.035em}
{\rm C\mkern-8mu\vrule height.65ex depth-.1ex
width.025em\mkern8mu\kern-.025em}}}

\def\RR{{\rm I\kern-1.6pt {\rm R}}}

\def\ZZ{{\rm Z}\kern-3.8pt {\rm Z} \kern2pt}
\def\IB{\relax{\rm I\kern-.18em B}}
\def\ID{\relax{\rm I\kern-.18em D}}
\def\II{\relax{\rm I\kern-.18em I}}
\def\IP{\relax{\rm I\kern-.18em P}}

  \def\tvf{\tilde{\varphi}}

  \def\th{\theta}                  

\def\tt{\tilde{\theta}}

\def\6{\partial}

\newfont{\namefont}{cmr10}
\newfont{\addfont}{cmti7 scaled 1440}
\newfont{\boldmathfont}{cmbx10}
\newfont{\headfontb}{cmbx10 scaled 1728}
\begin{document}
\setcounter{equation}{0}
\preprint{KUL-TF-08-10}
\preprint{ITP-UU-08/33}
\preprint{SPIN-08/24}

\title{Heavy quark potential with dynamical flavors: a first order transition}

\author{Francesco Bigazzi$\,^{a}$, Aldo L. Cotrone$\,^{b}$, Carlos N\'u\~nez$\,^{c}$, Angel Paredes$\,^{d}$}

\affiliation{\textit{a} Physique Th\'eorique et Math\'ematique and International Solvay
Institutes, Universit\'e Libre de Bruxelles; CP 231, B-1050
Bruxelles,
Belgium.\\
\textit{b}  Institute for theoretical physics, K.U. Leuven;
Celestijnenlaan 200D, B-3001 Leuven,
Belgium.\\
\textit{c} University of Wales Swansea, Dept. of Physics; Singleton Park, Swansea, SA2 8PP, Wales UK.\\
\textit{d} Institute for Theoretical Physics, Utrecht University; Leuvenlaan 4,
3584 CE Utrecht, The Netherlands.
}

\email{fbigazzi@ulb.ac.be, Aldo.Cotrone@fys.kuleuven.be, C.Nunez@swansea.ac.uk, A.ParedesGalan@uu.nl}

\begin{abstract}

We study the static potential between external quark-antiquark pairs in a strongly coupled gauge theory with a large number of colors and massive dynamical
flavors, using a dual string description.
When the constituent mass of the dynamical quarks is set below a certain critical
value, we find a first order phase transition between a linear
and a Coulomb-like regime. Above the critical mass the two phases are
smoothly connected.
We also study the dependence on the theory parameters of the quark-antiquark separation at which the static configuration decays into specific static-dynamical mesons.

\end{abstract}

\pacs{11.25.Tq,12.38.Aw,12.39.Pn,14.65.-q}
\maketitle

{\bf 1. Introduction}.-- The study of non-perturbative effects of dynamical quarks is a notoriously difficult problem to address with the present computational techniques in QCD.
Phenomenological models, such as the screened version \cite{lattice} of
the Cornell potential for heavy quarks \cite{cornell}, are of fundamental importance but an understanding of the physics of these effects from first principles is still missing.
In fact, the best tool for the study of QCD at strong coupling, i.e. its lattice formulation, is still partially limited by the lack of suitable computational power when dealing with dynamical light flavors (for our purposes the relevant reference is \cite{Bali:2005fu}).
In the latest years, string theory has developed as a possible tool for studying strong coupling effects in quantum field theory.
It is clearly not an ideal setting, since it allows to study mainly the large 't Hooft coupling, large $N_c$ (number of colors) limit of theories which typically include a number of other fields besides (possibly) the QCD ones.
Nevertheless, apart from its obvious theoretical interest, string theory has proven to be a valuable way for understanding processes which are universal enough, such as specific Quark-Gluon Plasma ones.

In this letter we study the ``connected part'' (i.e. we do not include the mixing with the meson/anti-meson states) of the static potential between two heavy test quarks in a SQCD-like quantum field theory, which includes light dynamical quarks of mass $m_q$, by using its dual string theory description.
Below a critical mass $m_c$ of the dynamical quarks, we observe a first 
order phase transition in the potential $V(L)$ as we increase the 
distance $L$ between the two test quarks.
The transition takes place in the region where the potential turns from Coulomb-like to linear.
Instead, for masses larger than $m_c$, the connection between the two 
regions is perfectly smooth.
We also study the ``string breaking length'' $L_{sb}$ at which the heavy quark configuration decays into a specific pair of heavy-light mesons.
It is shown that $L_{sb}$ is a decreasing function of the number $N_f$ and the mass $m_q$ of the dynamical flavors, the latter being a fully
non-perturbative effect.

While the theory we study is not QCD, it is possible that the main features of the static potential are sufficiently universal. Hence, we provide an analysis of non-perturbative effects of light dynamical quarks in a strongly coupled quantum field theory, in a very straightforward and simple way.

Sections 2 and 3 provide the technical details on the string set-up. Section 4 and 5, which can be read independently, contain the results. We define $x\equiv \frac{N_f}{N_c}$ and limit our analysis to the range $0<x<1.5$ where we reach sufficient numerical precision.

{\bf 2. Details on the dual string description}.--
The SQCD-like theory we consider is realized at the four dimensional
intersection of $N_c\gg1$  ``color'' D5-branes wrapped on a $S^2$ and
$N_f \sim N_c$ ``flavor'' D5-branes. It is a particular version of ${\cal N}=1$ SQCD deformed by a quartic
superpotential as considered in \cite{Casero:2006pt}. In these models, the flavor branes are homogeneously
smeared along the (large) internal manifold \cite{Bigazzi:2005md}. The typical separation
between two branes is much larger than the string scale so the flavor
symmetry is in fact $U(1)^{N_f}$. We
find solutions of IIB supergravity coupled to the flavor brane sources.
These sources are represented as a sum of DBI+WZ actions \cite{Klebanov:2004ya}. Notice that
even if the DBI action for $N_f$ branes is not valid when $g_s N_f$ is
not small, here we can use it because the branes are separated far apart so
what we have is
a sum of actions for individual branes. The string frame metric reads
\bear ds^2 &=& \alpha'e^\phi N_c \Big[ \frac{1}{\alpha'g_sN_c}dx_{1,3}^2 + 4 Y d\rho^2 \rc 
&& + H  (d\theta^2 +\sin^2 \theta d\varphi^2 ) + G  (d\tt^2 +\sin^2 \tt
d\tvf^2 ) \rc  && + \frac{a}{2} \cos \psi (d\theta d\tt -\sin\theta
\sin\tt d\varphi d\tvf)   \rc && + \frac{a}{2} \sin \psi (\sin\theta d\tt
d\varphi + \sin\tt d\theta d\tvf)   \rc  && + Y \left( d\psi + \cos\tt
d\tvf  +\cos \th  d\varphi  \right)^2 \Big].
\label{SDmetric} \eear
The dilaton $\phi$ and the functions $Y,H,G,a$ depend only on
the radial coordinate $\rho$; we used the notation $g_s\equiv e^{\phi_{IR}}$, where $\phi_{IR}\equiv\phi(\rho=0)$. In the massless flavor case, which was
studied in  \cite{Casero:2006pt}, the $N_f$ smeared branes were
extended along $\rho$ from the origin up to infinity. In the
following we will consider the case in which all the flavor branes
are extended up to a finite distance $\rho_q> 0$ from the origin.
This will correspond to turning on a mass for all the flavors in the
gauge theory dual.

The full supergravity background contains also a RR three-form given by
\bear  F &=& \alpha'g_sN_c \Bigl[-\frac{1}{4} \sin \tt d\tt \wedge d\tvf \wedge
\hat \omega_3 +\frac14 (d\theta \wedge d\tt  \rc &&  -
\sin \theta \sin\tt d\varphi \wedge d\tvf) \wedge (b \sin \psi \hat
\omega_3  - (\partial_\rho b) \cos \psi d\rho ) \rc && + \frac14
(-\sin\tt d\theta \wedge d\tvf + \sin \theta d\tt \wedge d\varphi)\wedge
(b \cos \psi \hat \omega_3 \rc && - (\partial_\rho b) \sin \psi
d\rho )- \frac{x -1}{4} \sin \theta d\theta \wedge
d\varphi \wedge \hat \omega_3 \Bigr],
\label{SD3form}\nonumber
\eear
where $\hat \omega_3 = d\psi
+\cos\tt d\tvf  +\cos \th  d\varphi$.
In  \cite{Casero:2006pt}, a set of differential equations
and constraints for $H,G,b,a,Y,\phi$ was found. They read, in the present notation,
\bear &&H = \frac{C ( a + b)}{4} - \frac{2 - x}{8},\qquad G = \frac{C( a - b) }{4} + \frac{2 - x}{8},\rc 
&& \partial_\rho b = - \frac{2 b C + x - 2}{S} ,\qquad \partial_\rho a = - \frac{2 a C + x - 8Y}{S} ,\rc
&& \partial_\rho Y = 8SY \frac{2 a^2 C -b (2 b C - 2 +x ) - a (x + 8 Y)}
{4a^2 S^2 - (2b C -2 + x)^2}, \rc
&& \partial_\rho \phi = -\frac12 \partial_\rho \log (4a^2 S^2 - (2b C -2 + x)^2) \rc && \qquad \quad +
\frac{32 a S^2 Y }{S
 (4a^2 S^2 - (2b C -2 + x)^2) },
\label{BPSsystem} 
\eear 
where $C = \cosh (2\rho),\ S=
\sinh (2 \rho)$. 
This system of equations is valid for arbitrary $N_f$, $N_c$.

\begin{figure}[htbp] 
 \centering
 \includegraphics[width=0.23\textwidth]{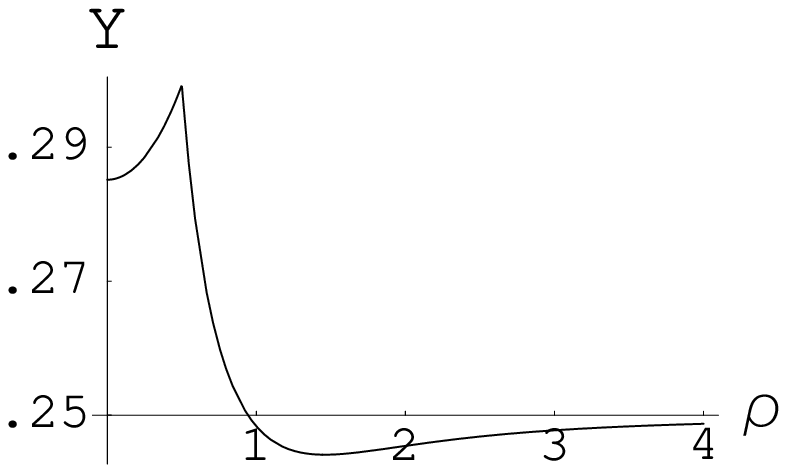}  \includegraphics[width=0.23\textwidth]{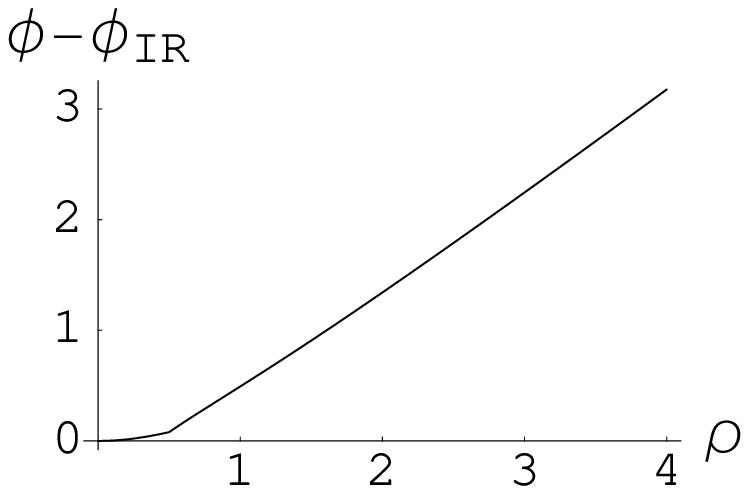}
 \caption{The $Y,\phi$ functions in the case $x = 0.8,\rho_q=0.5$.}
 \label{x1rhoM1}
 \end{figure}

The expressions (\ref{BPSsystem}) make the addition of massive flavors relatively easy. As it was noticed in a similar context
\cite{flavoredKW}, this can be effectively achieved by just
replacing the constant $x$ with a function of $\rho$, $x (\rho)$.
In this way the expression for the three-form flux
is unchanged, modulo the above redefinition of $x$. Indeed the
Bianchi identity gets changed, but since in the supersymmetric
variations for the background fermions only the field $F$ (and 
not its derivative $dF$)
enters, the whole BPS equations will not be modified in form. 
On the field theory side, we can neglect the quark masses at high
energies. At energies smaller than the quark masses, the quarks can
be integrated out and the theory looks like the unflavored one (a confining SYM-like theory in the present case). 
This suggests, via the holographic radius-energy relation, that we can
choose, as a model for $x(\rho)$, just a Heaviside step function
$x (\rho) = x \Theta (\rho - \rho_q)$.
At the
level of the supergravity equations of motion, the only constraint on
$x(\rho)$ is that $x(\rho)'\geq 0$ \cite{flavoredKW}. The exact form of the function 
should follow
from the smearing of the massive embedding associated to the flavor
branes.
Nevertheless, the Heaviside approximation is expected to
correctly capture
the qualitative features of the system under study, as it also happens
in other similar setups \cite{paper2}.
Since the first two equations in (\ref{BPSsystem}) are constraints
and not differential equations, not all
the functions can be continuous if $x(\rho)$ is discontinuous. We 
impose that the metric is continuous so $H,G,a,Y,\phi$ are
continuous at $\rho = \rho_q$. 
This is the equivalent procedure to matching the couplings
when one integrates out quarks in the field theory.
The only
function which does not enter the metric is $b$. 
In any case, what should be continuous - see eq. (\ref{BPSsystem}) - is
$\frac{C}{4}b + \frac{x}{8}$. The solution for $b$ consistent with
the differential equations and this condition is: \bear b = \frac{2
\rho}{\sinh (2\rho)}  + x\,\Theta (\rho - \rho_q) \left(
\frac{\rho_q-\rho - \frac{\sinh (2\rho_q)}{2 \cosh
(2\rho_q)}}{\sinh(2\rho)}\right). \label{bmassive} \nonumber \eear Below
$\rho_q$, we have the unflavored system, so we can take
the regular solution, found in Section 8 of the first paper in \cite{Casero:2006pt}, depending 
on a parameter $\mu$. 
Tuning $\mu$ for a given pair $(x$, $\rho_q)$, it is possible to get a
solution with the expected UV which is uniquely determined up to the additive
constant $\phi_{IR}$.
The system can be solved numerically. The plots of figure
\ref{x1rhoM1} depict a typical case for the functions $Y,\phi$.

{\bf 3. The quark-antiquark potential}.-- 
We focus on the interaction energy between two test quarks $Q,\bar
Q$ put as external probes of our theory. The string dual of the
$Q{\bar Q}$ system is a macroscopic open string attached to a probe
flavor brane which is very far from the bottom of the space ($\rho_Q \gg \rho_q$), so that
the associated quark mass
$M_Q\rightarrow\infty$ \cite{Rey:1998ik}. The string bends in the bulk and reaches a
minimal radial position $\rho_{0}$. The Minkowski separation $L$
between the test quarks, as well as the total energy
of the system, depends on $\rho_{0}$. From this we can deduce the
$V(L)$ relation, where $V(L)$ is the potential (renormalized energy), i.e. the
total energy minus the total quark mass $2M_Q$. The open string embedding is chosen as $t=\tau,
y=\sigma, \rho=\rho(y)$ where $y\in [-L/2,L/2]$ is one of the
spatial Minkowski directions. The string worldsheet action reads \bear
S= -\frac{1}{2\pi\alpha'}\int dt dy \sqrt{g_{tt}(\rho)[g_{yy}(\rho)
+ (\partial_y \rho)^2 g_{\rho\rho}(\rho)]}. \nonumber \eear 
Defining
$f \equiv \sqrt{g_{tt}g_{yy}}=e^{\phi-\phi_{IR}},\, g \equiv \sqrt{g_{tt}g_{\rho\rho}}=2e^{\phi-\phi_{IR}}\sqrt{Y}\sqrt{g_sN_c\alpha'}$, 
the (constituent) quark mass is given by the energy of a string 
extended from the bottom of the space at $\rho=0$ to the bottom
of the corresponding flavor brane, $m_q=\frac{1}{2\pi\alpha'}\int_{0}^{\rho_{q}} g \ d\rho$, $M_Q=\frac{1}{2\pi\alpha'}\int_{0}^{\rho_{Q}} g \ d\rho$. 
We write the
length $L$ and potential $V$ as functions of 
$\rho_0$ and with UV cutoff $\rho_{Q}$
as (the $0$ subindex means that
the quantity is evaluated at $\rho=\rho_0$): \bear L(\rho_0)&=&2
\int_{\rho_0}^{\rho_{Q}} \frac{g f_{0}}{f \sqrt{f^2
-f_{0}^2}}d\rho\,\,,\rc V (\rho_0)&=&\frac{2}{2\pi\alpha'} \Bigl[\int_{\rho_0}^{\rho_{Q}}
\frac{g f}{\sqrt{f^2 -f_{0}^2}}d\rho -
\int_{0}^{\rho_{Q}} g \ d\rho \Bigr]. \nonumber \eear It is not difficult
to check the relation $\frac{dV}{d\rho_0} = \frac{f_{0}}{2\pi\alpha'}
\frac{dL}{d\rho_0}$. This for instance means that
$\frac{dV}{d\rho_0}$ and $\frac{dL}{d\rho_0}$ have the same sign.
In some cases they can change sign simultaneously at some
value of $\rho_0$ so the $V(L)$ plot turns around. 

We can now plug our numeric solutions into the definitions of the
 quark-antiquark potential $V(\rho_0)$ and distance $L(\rho_0)$, to see how they
depend on the dynamical massive flavors. In doing so we
consider a finite (large) UV cutoff, and so a finite $M_Q\gg
m_q$, which 
we keep fixed as $m_q$ and $x$ are
varied (we will show the results for $M_Q=200\sqrt{g_sN_c/\alpha'}$ but their qualitative features do not depend on this choice).

\begin{figure}
 \centering
\includegraphics{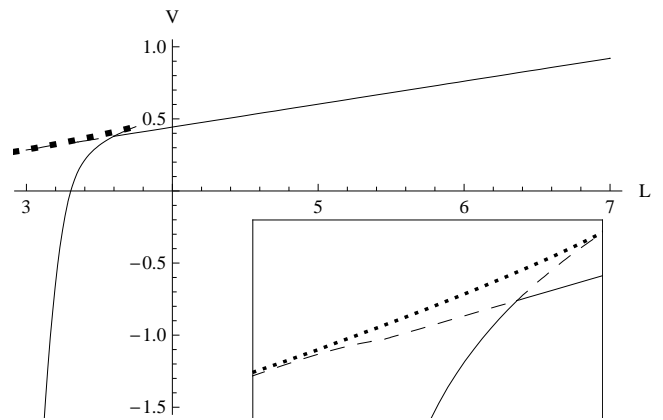}
\caption{The potential for small dynamical mass $m_q=0.01$ and $x=0.8$. In the frame, a zoom of the critical region. Dashed (dotted) lines are metastable (unstable); cfr. \cite{Avramis} for stability issues. Lengths are in units of $\sqrt{\alpha'g_sN_c}\sim 1/\Lambda_{QCD} $, energies in units of $\sqrt{\alpha'g_sN_c}/\alpha'\sim T/\Lambda_{QCD}$, $T$ being the string tension.}\label{gloriousturnaround}
\end{figure}

{\bf 4. A first order transition in the potential}.--
In the $m_q=0$ case, the $\bar Q Q$ string breaks at a certain
$L_{max}$, where the $V(L)$ plot turns around \cite{Casero:2006pt}. At $m_q\neq0$, this turn-around (that one would expect at least in the small mass limit), should be followed by a second turn-around, such that the potential comes back to an increasing linear behavior at large $L$ (i.e. in the far IR, where the quarks are integrated out and the theory is SYM-like). Our numerical analysis confirms this expectation, adding a crucial piece of information: the existence of a critical value $m_c$ for the dynamical quark mass. For $m_q>m_c$, the $V(L)$
plot is monotonic, a Coulomb-like behavior at small $L$ smoothly joining with the linear one for large $L$.
Instead, in the small mass regime, $m_q<m_c$,
we find a first order \footnote{We are grateful to Marco Caldarelli for pointing this to us.} transition between the Coulomb-like and the linear phase, with a corresponding double turn-around behavior of $V(L)$ at intermediate distances, see figure \ref{gloriousturnaround}. In the figure,
\begin{figure}
 \centering
\includegraphics{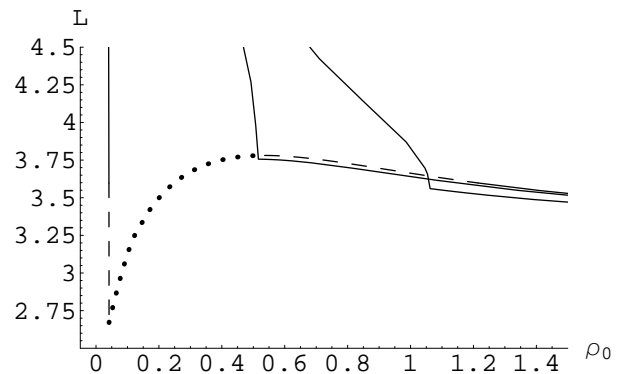}
\caption{Some representative $L(\rho_0)$ curves at $x=0.8$ and $m_q=0.01,0.09,0.2$ (left to right). Solid, dashed and dotted lines correspond to the regions of figure \ref{gloriousturnaround}.}\label{turnaround}
\end{figure}
the energetically favored branch of the graph is the one with largest $L$ at a given $V$ (the solid line). 
So, there is a discontinuity in the derivative of $V(L)$, with a sudden decrease of the ``local string tension'', such that at a certain energy
scale and length the increase of the energy of the
flux tube with the length has a deceleration. The Coulomb-like regime does not connect smoothly with the linear regime \footnote{One may think that the behavior in Fig \ref{gloriousturnaround} is due to a level crossing with a hybrid potential. We than Joan Soto for a discussion on this point.}. 

The existence of a critical mass $m_c$ above which there is no phase transition, is evident by analyzing the $L(\rho_0)$ curve at different values of
$m_q$. 
For example, for x=0.8, one can check from plots as those in figure
\ref{turnaround} that
below $m_c \sim 0.09$, there is a first order phase transition between a
large $\rho_0$
string and a small $\rho_0$ one. Actually these
plots share some qualitative features with the pressure/volume isothermal curves in a van der Waals gas, where the phase transition,
occurring below a certain critical temperature, is between a
liquid and a gas phase. 
It might prove useful to model the ``connected part'' of the phenomenological
potential between
two heavy quarks with some function, similar to the Gibbs free energy/pressure
relation in the
van der Waals system, exhibiting a discontinuity in the first derivative
for small dynamical quark mass. We checked that the quark distance at which the transition occurs is a decreasing function
of  $x$ and an increasing function of $m_q$. On the other hand, the value of
the potential at the transition is a decreasing function of both $x, m_q$. Finally, $m_c\rightarrow 0$ for $x\rightarrow 0$, where one recovers the quenched approximation.

{\bf 5. String breaking length}.--
In the presence of dynamical quarks there is a screening length, 
defined as the length at which the $\bar Q Q$ string will break by production of quark pairs.
It is determined by the condition that the energy of the Wilson loop equals the energy of two mesons
composed by one test and one dynamical quark.
In the case at hand, due to the smearing procedure, the lighter of such mesons are nearly massless.
In fact, in the string language they are localized at the intersection of the test brane and the dynamical brane.
Their mass should scale as $M_Q/(g_s N_c)$, with $g_sN_c\gg1$ \cite{Herzog:2008bp}.
This is parametrically smaller than the typical energy $E_{{\bar Q}Q} (L)=2M_Q+V(L)$ computed above.
Thus, the Wilson loop has always enough energy to decay into these mesons and the screening length is not a relevant quantity.
Nevertheless, again due to the smearing, the decay in this channel (and in every channel corresponding to 
a specific flavor) is $1/N_c$ suppressed.
If we ask for a decay rate which is not $1/N_c$ suppressed we have to consider a larger $\bar Q,Q$ separation, such that 
the $\bar Q Q$ state can decay into a sizable fraction of the $N_f$ types of heavy-light mesons.
Given such a fraction, one could study how the separation $L_{sb}$, required for the decay on all the included flavors to happen, depends on the parameters $m_q,x$. 
Clearly, the choice of the fraction would be arbitrary.

In order to get an idea of the general trend, we choose to define $L_{sb}$ as the length at which the Wilson loop has 
enough energy to decay by production of the particular flavor which has precisely the same charge (internal or R-symmetry quantum numbers) as the test quarks.
The string dual to the corresponding specific meson, whose mass is $E_{\bar q Q}=M_Q-m_q$, is stretched only along the radial direction of the geometry (\ref{SDmetric}), from the bottom of the test brane to that of the ``parallel'' dynamical one \cite{Karch:2002xe}.
We call $L_{sb}$ the ``string breaking length''. The value $L_{sb}$ at which the breaking can occur
is given by equating the energies of the initial and final configurations: $E_{{\bar Q}Q} (L) - E_{{\bar q}Q + q {\bar Q}}=V(L) + 2m_{q}=0$. We are
neglecting the interactions between the two final mesons because their contribution is subleading due to the smearing.

\begin{figure}
\centering
\includegraphics[width=0.23\textwidth]{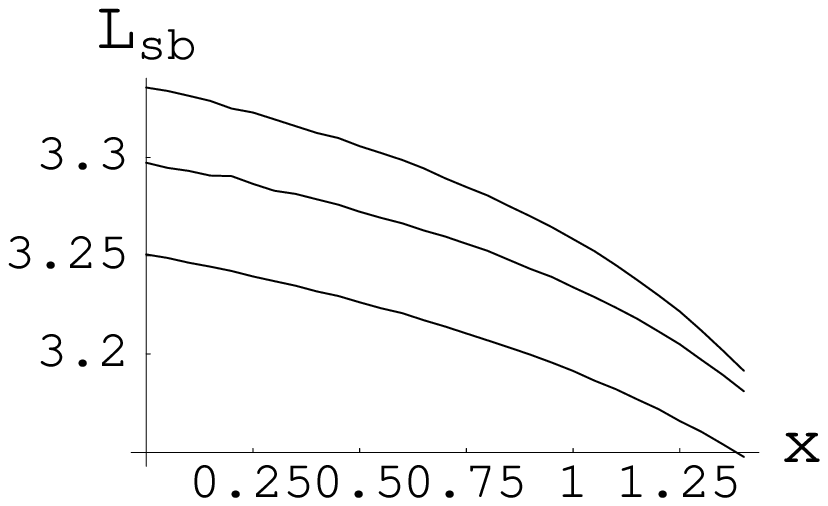} \includegraphics[width=0.23\textwidth]{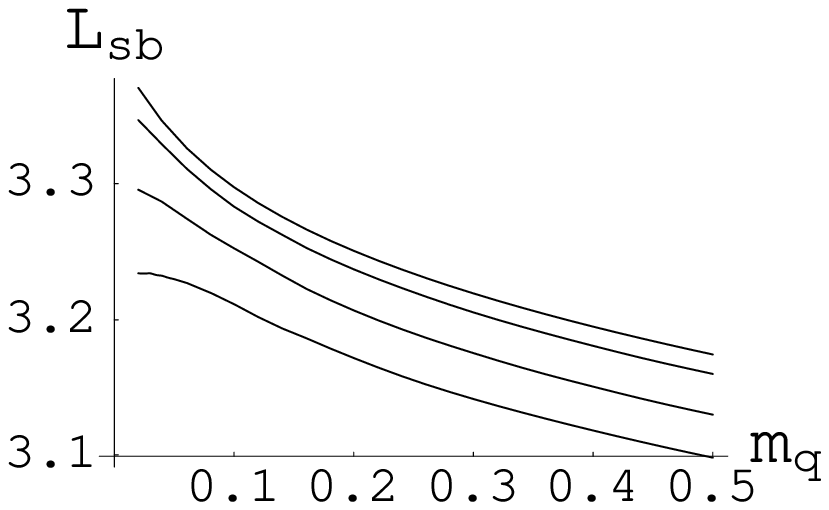}
\caption{The string breaking length $L_{sb}$ as a function of: (left) the number of flavors $x$ for $m_q=0.05, 0.1, 0.2$ (bottom to top) and (right) the mass $m_q$ for $x=0, 0.3, 0.8, 1.2$ (top to bottom).}\label{turnaround1}
\end{figure}

The results are in figure \ref{turnaround1}. 
The string breaking length is a decreasing function of the number of flavors $x$: the more flavors there are in the theory, the easiest is the decay.  Finally, the string breaking length is a decreasing function of $m_q$;
this is a genuine large coupling effect, and may change if one chooses a different
definition of $E_{\bar q Q}$.

{\bf Acknowledgments}.-- It is a pleasure to thank Roberto Casero, who participated in the early stages of this project, and Marco Caldarelli for a crucial observation on the phase transition. We also thank R. Argurio, A. Di Giacomo, N. Drukker, F. Ferrari, S. Hands, E. Imeroni, B. Lucini, A. Rago, J. Soto, M. Teper, W. Troost.
This work is supported by the EU contract MRTN-CT-2004-005104, by the FWO -
Vlaanderen project G.0235.05, by the Federal Office for
STC Affairs programme P6/11-P, by the Belgian
FRFC (G.2.4655.07), IISN (G.4.4505.86) and
the IAPP (Belgian Science Policy).
AP is also supported by a NWO VIDI grant 016.069.313 and by
INTAS contract 03-51-6346. 

\end{document}